\documentstyle[12pt]{article}
\def\tib{\tilde{b}}
\def\tic{\tilde{c}}
\def\tid{\tilde{d}}
\def\tig{\tilde{g}}

\def\aw{$a_\mu^w$}
\def\bw{$b_\mu^w$}
\def\cw{$c_{\mu\nu}^w$}
\def\dw{$d_{\mu\nu}^w$}
\def\ew{$e_\mu^w$}
\def\fw{$f_\mu^w$}
\def\gw{$g_{\la\mu\nu}^w$}
\def\Hw{$H_{\mu\nu}^w$}

\def\al{\alpha}
\def\be{\beta}
\def\ze{\zeta}
\def\et{\eta}
\def\la{\lambda}
\def\om{\omega}
\def\De{\Delta}
\def\Om{\Omega}

\def\fr#1#2{{{#1} \over {#2}}}

\def\X{{\hat X}}
\def\Y{{\hat Y}}
\def\Z{{\hat Z}}

\def\bea{\begin{eqnarray}}
\def\eea{\end{eqnarray}}
\def\ket#1{|{#1}\rangle}
\def\etal{{\it et al.}}

\newcommand{\beq}{\begin{equation}}
\newcommand{\eeq}{\end{equation}}

\renewenvironment{thebibliography}[1]
 { \rm
   \begin{list}{\arabic{enumi}.}
    {\usecounter{enumi} \setlength{\parsep}{0pt}
     \setlength{\itemsep}{3pt} \settowidth{\labelwidth}{#1.}
     \sloppy
    }}{\end{list}}

\begin{document}

\begin{center}
{{\large\bf Space-Based Searches for Lorentz and CPT Violation
\footnote{From the 2002 NASA/JPL Workshop
for Fundamental Physics in Space, Dana Point,
CA, May 2002.}\\} \vspace{0.2cm}
Neil Russell\\
{\small\it Physics Department, Northern Michigan University,
Marquette, MI 49855, U.S.A.\\
email: nrussell@nmu.edu\\}}
\vspace{0.3cm}
\parbox{6in}{\small
In this proceedings,
a summary is presented of recent research investigating ways in
which high-precision atomic clocks on the International Space Station
could search for violations of Lorentz and CPT symmetry.
Space-based searches offer certain experimental advantages over
Earth-based experiments investigating these symmetries.
The results are based on work published in Physical Review Letters,
volume 82, article 090801, 2002.}
\end{center}

\vspace{.4cm}\noindent
{\bf 1. Introduction}\\[.2cm]
This contribution to the proceedings of the 2002 NASA/JPL Workshop
on Fundamental Physics in Space summarizes recent research
\cite{spaceexpt} aimed at using atomic clocks and other apparatus
on the International Space Station to search for violations
of Lorentz and CPT symmetry
at the Planck scale.
We consider generalities relating to experiments
mounted on spacecraft and
consider some tests that could be performed using
clocks planned for installation on the
International Space Station (ISS).
This work was done in collaboration
with Robert Bluhm, Alan Kosteleck\'y,
and Charles Lane.

Lorentz symmetry is a feature of
the standard model of particle physics.
A considerable body of research exists
investigating the possible violation of Lorentz
symmetry, however.
From the theoretical view,
the motivation for this effort lies in
discovering new physics beyond the standard model.
From the experimental side, the rapidly improving
sensitivities of various experiments
may reveal previously unresolved effects.
Recent theoretical work on Lorentz and CPT symmetry
includes the development of a framework
that allows for general minuscule violations
of these symmetries in the context of particle physics.
This framework is known as the
standard-model extension~\cite{ck}.

Associated with the standard-model extension is
a range of literature discussing a variety of theoretical issues,
as well as a growing number of experimental results
bounding possible effects.
The violation of Lorentz symmetry~\cite{cpt01}
may arise in the context of string theory,
and may be accompanied also by CPT violation~\cite{kps}.
Violation of Lorentz and CPT symmetry
has also been discussed in the context of
supersymmetry~\cite{bek},
and noncommutative field theory~\cite{chklo}.
The standard-model extension is expected to be
the low-energy limit of some fundamental underlying theory,
and so the violations would most likely be suppressed
by ratios involving the low-energy mass and the
10$^{19}$-GeV Planck mass.
The broad applicability of the standard-model extension
to all areas of physics is an attractive feature.
Among the interesting implications
is a possible mechanism for generating
the baryon asymmetry in the universe~\cite{bckp}.
For the neutral mesons,
some bounds on standard-model extension parameters
exist for the neutral K and D mesons,
and results are anticipated for the neutral B
system~\cite{kexpt,ak,bexpt,ckpvi}.
The symmetry properties of these meson systems
have interesting analogue models
in classical mechanics~\cite{kr}.
In the photon sector,
data from distant cosmological sources places
stringent bounds on Lorentz symmetry~\cite{ck,cfj,km}.
In the lepton sector,
recent results have come from a muonium experiment,
and from anomaly-frequency comparisons
of oppositely-charged muons
at CERN and BNL~\cite{vh}.
Earlier work considered electron-positron
comparisons using Penning traps~\cite{eexpt}.
Impressive results are possible with
a spin-polarized torsion pendulum~\cite{eexpt2}.

Of particular relevance here
are clock-comparison experiments with atoms
and ions~\cite{ccexpt,lh,db,dp}.
Such experiments can identify spectral lines with resolutions
at the Planck scale~\cite{kla}.
The general principle of a clock-comparison experiment
is to search for violations of rotational symmetry by
monitoring the frequency variations of
a Zeeman hyperfine transition as the quantization axis
changes direction.
Usually, the frequencies of two different clocks are monitored
as the laboratory rotates with the Earth.
To avoid issues with signals travelling between
two different locations, the clocks are co-located.
Placing such an experiment in a satellite may produce results
slightly better than have been achieved on earth,
and this proceedings aims to consider some of the issues
associated with this possibility.

\vspace{.4cm}\noindent
{\bf 2. Clocks and Inertial Frames}\\[.2cm]
Atomic transitions can be measured with great precision
and so are suitable candidates for time standards.
In conventional physics with constant laboratory conditions,
these clock frequencies are constant quantities.
However, in the standard-model extension
with Lorentz and CPT violation,
some Zeeman hyperfine transitions are
shifted in frequency~\cite{kla}.
For an experiment operating on such a transition,
these shifts are controlled at leading order
by parameters denoted in the clock reference frame as
$\tib_3^w$, $\tic_q^w$, $\tid_3^w$, $\tig_d^w$, $\tig_q^w$.
Here, the superscript $w$ is $p$ for the proton,
$n$ for the neutron, and $e$ for the electron.
These quantities are particular combinations
of the basic coefficients
\aw, \bw, \cw, \dw, \ew, \fw, \gw, \Hw\
appearing in the standard-model extension,
and are related to expectation values in
the underlying fundamental theory.
For example,
\beq
\tib_3^w = b_3^w -m_w d_{30}^w + m_w g_{120}^w -H_{12}^w \ ,
\eeq
where $m_w$ is the mass of the particle of type $w$
and the subscripts are indices defined in a reference frame
with the $3$ direction defined as the clock quantization axis.

In the case of an Earth-based laboratory,
the parameters $\tib_3^w$, $\tic_q^w$,
$\tid_3^w$, $\tig_d^w$, $\tig_q^w$
are not fixed, but vary in time
due to the sidereal rotation of the Earth
with period
$ 23\, {\rm h} \, 56 \, {\rm min} \simeq 2\pi/ \Om $.
The mathematical form of this time dependence
can be found by considering the transformation
from the laboratory frame containing the clock,
with coordinates numbered $(0,1,2,3)$,
to a suitable nonrotating frame with coordinates
$(T,X,Y,Z)$.
Ideally, an inertial nonrotating frame is required,
but for practical purposes any frame sufficiently inertial
for the desired experimental sensitivity may be selected.
Frames associated with the Earth, the Sun, the Milky Way galaxy,
or the cosmic microwave background radiation would be
possible choices for the inertial frame.

In earlier literature,
the nonrelativistic transformation from the clock frame to
the nonrotating frame has been considered
\cite{kla}.
In the case of space-based experiments,
leading-order relativistic effects are of interest.
An Earth-centered choice of reference frame must then
be rejected for such relativistic investigations
because it is inertial over a limited
time scale of perhaps a few days.
Frames centered on the Sun, the galaxy,
or the microwave background are approximately inertial
over thousands of years,
and are all acceptable for experiments.
The choice of frame must be stated
when reporting bounds on components of
coefficients of Lorentz violation,
since the numerical values will be frame-dependent.

A good choice of reference frame for our purposes
is one centered on the Sun.
So, we select the spatial origin on the Sun,
the $\Z$ unit vector parallel to the Earth's rotation axis,
the $\X$ unit vector in the equatorial plane
pointing at the celestial vernal equinox,
and $\Y$ completing the right-handed system.
The origin of the time variable $T$ is taken to be the
vernal equinox in the year 2000, using a clock located
at the spatial origin.
In this system,
the Earth orbits about the Sun
in a plane tilted at an angle of $\et \simeq 23^\circ$
relative to the $XY$ plane.

An adequate geometrical description of the orbital
configuration can be obtained by
approximating the Earth's orbit as
a circular trajectory with angular frequency $\Om_\oplus$
and speed $\be_\oplus$.
In addition, a satellite orbit about the Earth
is approximated as circular with angular frequency
$\om_s$ and speed $\be_s$.
We use $\ze$ to denote the angle between
$\Z$ and the axis of the satellite orbit.
We denote by $\al$ the right ascension angle of
the ascending node of the orbit.
In the case of the ISS,
$\al$ precesses by a few degrees per day.

Time intervals on a clock in a satellite
are dilated when seen from the inertial Sun frame.
Relative to the Sun-based frame,
the clock velocity is $\vec V(T)=d\vec X/dT$,
where the position vector $\vec X(T)$
of the clock is determined by positions of the
Earth and the spacecraft.
This vector $\vec V(T)$ is needed to obtain
an accurate conversion between the times in
the laboratory and in the Sun frame.
In principle,
effects such as perturbations in this vector
and in the gravitational potential
should be included in this description.
In practice,
these corrections may be neglected because
the experiments involve comparing two clocks
within the same satellite,
which are essentially at the same location.
In this case, standard relativity predicts
identical rates of advance of the clocks.
However,
in the presence of Lorentz and CPT violation,
clocks composed of different atomic species
will be differently affected,
despite being co-located .

Pertinent issues exist concerning the optimal
orientation of the clock quantization axis
relative to the geometric configuration of the system.
If the clock apparatus is fixed within the satellite,
the flight mode of the satellite
will determine the clock quantization axis
relative to the Sun frame.
For this proceedings,
we focus on a flight mode
with quantization axis tangential to
the circular satellite trajectory about the Earth.
We choose the clock reference frame with
3 axis parallel to the satellite motion about the Earth,
1 axis pointing towards the center of the Earth,
and 2 axis perpendicular to the satellite orbital plane.
This configuration would be possible with some clock experiments
on the ISS.
The results outlined here are specific examples,
but we note that other modes of flight and quantization-axis
configurations can be handled by the methods discussed here.
It is important to note that
sensitivity to some components
is only possible with specific quantization-axis orientations.

Experiments searching for Lorentz and CPT violation in the
context of the standard-model extension
are aimed at measuring the tensor-like
parameters \aw, \bw, \cw, \dw, \ew, \fw, \gw, \Hw\
in our standard solar reference frame.
Measurements made in the laboratory
frame must be transformed to the Sun-based frame
by taking into account the relevant rotation and boost
$\vec V(T)$.
This means that the components
of the coefficients for Lorentz violation
in the clock frame must be expressed
in terms of components in the Sun-based frame.
To give an example,
the transformation of the component $b_3^w$ is
\bea
b_3^w &=& b_T^w \{\be_s -
\be_\oplus [\sin \Om_\oplus T (\cos \al \sin \om_s \De T
  \nonumber \\
&& \qquad \qquad + \cos \ze \sin \al \cos \om_s \De T ) - \cos \et
\cos \Om_\oplus T
  \nonumber \\
&& \qquad \qquad \times (\sin \al \sin \om_s \De T - \cos \ze \cos
\al \cos \om_s \De T )
  \nonumber \\
&& \qquad \qquad + \sin \et \cos \Om_\oplus T \sin \ze \cos \om_s
\De T]\}
  \nonumber \\
&& - b_X^w (\cos \al \sin \om_s \De T + \cos \ze \sin \al \cos
\om_s \De T )
  \nonumber \\
&& - b_Y^w (\sin \al \sin \om_s \De T - \cos \ze \cos \al \cos
\om_s \De T )
  \nonumber \\
&& +b_Z^w \sin \ze \cos \om_s \De T, \label{b3}
\eea
where $\De T = T - T_0$
is the time interval measured
from an agreed reference time $T_0$.
This transformation ignores effects such as
the Thomas precession,
holding only up to leading order in the velocities.
The above result for $b_3^w$
has to be included with the transformations for
the other coefficients to get the full result for
the observable parameter $\tib_3^w$ in the Sun frame.
The other coefficients $\tic_q^w$, $\tid_3^w$, $\tig_d^w$,
and $\tig_q^w$ are found by a similar method.
The expressions that result depend on combinations
of basic coefficients for Lorentz and CPT violation,
on trigonometric functions of various angles,
on frequency-time products,
on $\be_\oplus$,
and on $\be_s$.

\vspace{.4cm}\noindent
{\bf 3. Signal Features}\\[.2cm]
Satellite-based experiments offer accessibility to
all the spatial components of the basic coefficients for Lorentz
and CPT violation.
This eliminates a major constraint due to the fixed rotation axis
for Earth-based experiments,
preventing sensitivity to various spatial components.
For instance, ground-based experiments
sensitive to the laboratory-frame parameter $\tib_3^w$
would in turn be sensitive only to
the nonrotating-frame components
$\tib_X^w$, $\tib_Y^w$.
They can therefore bound only a limited subset of
components of \bw, \dw, \gw, \Hw.
This limitation would be overcome by
a satellite platform.
In the case of most satellites, the orbital axis is tilted
relative to the Earth's rotation axis,
and the orientation of this orbital axis precesses about the steady
axis of the Earth.
This precession makes the other spatial directions accessible
to satellite tests.

Another attractive feature of the satellite platform for experiments
is the relatively short orbital period.
Since the satellite orbital period $2 \pi /\om_s$ for
low-altitude satellites
is much less than a sidereal day, data can be collected in a
substantially reduced period.
In the case of the ISS, the 92-minute orbital period translates into
a data-collection period approximately 16 times shorter
than on Earth, where the orbital period is about 24 hours.
This could contribute to better results since it would reduce the
sensitivity loss due to clock instabilities over time.
One interesting advantage of this reduced experimental time
is due to the fact that the Earth's velocity vector
would remain essentially constant over the experimental duration.
This makes it possible to analyze the
leading-order relativistic effects due to the speed $\be_\oplus
\simeq 1\times 10^{-4}$ of the Earth relative to the Sun.
Such tests are not possible with ground-based experiments,
because they require several months of data,
during which time the velocity of the Earth changes significantly.
The analysis would be considerably simplified by the fact that
the Earth could be regarded as an inertial reference frame.
Direct extraction of leading-order relativistic effects
would be possible.

The observations above show that many types of Lorentz
and CPT violation that are unconstrained to date
would be accessible in space-based experiments.
As an example, consider a clock-comparison experiment
with sensitivity to the observable $\tib_3^w$
for particle species $w$.
In the Sun-based frame and for each $w$,
this observable is a linear combination of the basic coefficients
\bw, \dw, \gw, \Hw\ for Lorentz violation,
numbering 35 independent observable components if
the effect of field redefinitions is allowed for.
Whereas a conventional ground-based experiment is
sensitive to 8 of these,
the same type of experiment mounted on a space platform would
be sensitive to all 35.
Another approach to overcoming constraints on
accessible coefficients would be to construct
a suitably-oriented rotating base for a ground-based experiment.
This option is not pursued here, since the current work is aimed at
understanding sensitivities of experiments planned for the ISS.

For ground-based experiments,
some relativistic Lorentz and CPT coefficients
are suppressed by the boost factor of the Earth, $\be_\oplus$.
In comparison, space-based clock-comparison experiments
would also be sensitive to
first-order relativistic effects proportional to
the boost factor of the satellite, $\be_s$.
In Earth-based experiments, investigating the corresponding effects
of the lab motion relative to the Earth's center
would be impractical.
Such effects would also be further
suppressed by $\Om/\om_s$,
which is about $6 \times 10^{-2}$
in the case of the ISS.

A somewhat unexpected effect exists among the order-$\be_s$
corrections.
It is found that in space-based experiments
a dipole shift can lead to a potentially detectable signal
with frequency $2\om_s$.
This is not seen in the nonrelativistic analysis of
Earth-based clock-comparison experiments,
where signals with the double frequency $2\Om$ occur only
for quadrupole shifts.
To better understand this,
consider the parameter $\tib_3^w$, which
nonrelativistically is the third component of a vector and
would lead only to a signal with frequency $\om_s$.
This parameter $\tib_3^w$ contains the component $d_{03}$,
however, which in a relativistic approach
behaves like a two-tensor at leading order in $\be_s$,
and would therefore lead to a signal at frequency
$2\om_s$.
We give an example:
when the Earth is near the
northern-summer solstice,
$\tib_3^w$ in the Sun-based frame
has a double-frequency term that goes like
$\cos (2 \om_s \De T)$ with coefficient $C_2$ containing
the following spatial components of \dw:
\bea C_2 &\supset& \be_s \fr m 8 [ \cos
2 \al (3+\cos 2 \ze) (d^w_{XX} - d^w_{YY})
\nonumber \\
&& + (1- \cos 2 \ze) (d^w_{XX} + d^w_{YY} - 2 d^w_{ZZ})
\nonumber \\
&& - 2 \sin 2 \ze (\cos \al \, (d^w_{YZ} + d^w_{ZY}) - \sin \al \,
(d^w_{ZX} + d^w_{XZ}))
\nonumber \\
&&
+ (3+\cos 2 \ze)\sin 2 \al \, (d^w_{XY} + d^w_{YX})] .
\label{2oms}
\eea
This shows that all observable spatial components of \dw
could be accessed through appropriate monitoring
of the $2\om$ frequency.

\vspace{.4cm}\noindent
{\bf 3. Experiments on Earth Satellites}\\[.2cm]
The ISS will house a number of high-precision clocks and
other oscillators capable of testing fundamental physics
in the coming years.
Instruments slated for installation include H masers,
laser-cooled Cs and Rb clocks, and
superconducting microwave cavity
oscillators~\cite{parcs,aces,race,sumo}.
Among the experimental advantages of the ISS are
the orbital parameters
$\be_s \simeq 3 \times 10^{-5}$ and $\ze \simeq 52^\circ$,
which correspond to a speed and orbital plane
outside the scope of Earth-based experiments.
In addition, experiments on the ISS
would be conducted in
a microgravity environment with
reduced environmental disturbances,
and these features are expected to lead to sensitivity gains
compared with ground-based clocks.
The analysis presented in this proceedings is valid for
tests with all these clocks,
but not for the oscillators,
which are discussed elsewhere~\cite{km}.

In our discussion,
we consider a canonical configuration
with a signal clock being compared to
a co-located reference clock.
The signal clock is sensitive to
leading-order Lorentz and CPT violation,
while the reference clock,
for example an H maser tuned to its clock transition
$\ket{1,0} \rightarrow \ket{0,0}$,
is insensitive to such effects.

\vspace{.4cm}\noindent
{\bf Hydrogen Masers}\\[.2cm]
A hydrogen maser operating on the
transition $\ket{1,\pm 1} \rightarrow \ket{1,0}$
would be one possible signal clock.
A recent ground-based experiment used a
double-resonance technique to monitor this
transition frequency~\cite{dp},
which is sensitive to the
parameters $\tib_3^p$ and $\tib_3^e$ in the clock frame.
The sensitivity to relatively clean parameter combinations
is a consequence of the simplicity of the hydrogen system
as compared with atoms such as Rb or Cs used
in atomic clocks.
Mounting this experiment on the ISS would mean
that an experimental run of only about a day would suffice to
obtain data roughly equivalent to four months of data taken
on Earth with a similar experiment on a fixed base.
For both $w=e$ and $w=p$, all spatial components of \bw,
$m_w$\dw, $m_w$\gw, \Hw\ could be sampled by exploiting the orbital
inclination ($\ze\neq 0$) and by repeating the experiment
at a later time when orbital precession corresponds to a
significantly different value of $\al$.
Making the assumption of a 500 $\mu$Hz sensitivity,
equalling that attained in Earth-based experiments,
several presently unbounded components
would be probed at the level of about $10^{-27}$ GeV,
and others at about $10^{-23}$ GeV.
We also estimate that
cleaner bounds on certain spatial components of
$m_w$\dw, $m_w$\gw at the level of about $10^{-23}$ GeV
could be obtained by searching for a signal at the
double frequency $2\om_s$.
In all, about 50 components of coefficients
for Lorentz and CPT violation that are
currently unbounded could be tested at the Planck scale.

\vspace{.4cm}\noindent
{\bf Cesium Clocks}\\[.2cm]
In the case of a laser-cooled $^{133}$Cs clock,
a reference frequency could be provided by the usual clock transition
$\ket{4,0} \rightarrow \ket{3,0}$,
which is insensitive to Lorentz and CPT violation.
A Zeeman hyperfine transition such as $\ket{4,4} \rightarrow \ket{4,3}$
would be needed to provide a signal.
Since $^{133}$Cs has an unpaired electron,
this atom has sensitivity to electron parameters
similar to that of the H maser.
In the Schmidt model, the $^{133}$Cs
nucleus is a proton with angular momentum 7/2, giving
sensitivity to all clock-frame parameters $\tib_3^p$, $\tic_q^p$,
$\tid_3^p$, $\tig_d^p$, $\tig_q^p$, and yielding both dipole
and quadrupole shifts.
We note that components tested would include
$c_{\mu\nu}^p$.
Repeating results achieved in an Earth-based experiment would
imply a sensitivity level of about 50 $\mu$Hz~\cite{lh}
on the $\ket{4,4} \rightarrow \ket{4,3}$ transition.
A similar experiment on the ISS would potentially
run for a period reduced by a factor of 16.
Furthermore, measurements of the double-frequency signal $2\om_s$
would probe the spatial components of
$c_{\mu\nu}^p$ at the $10^{-25}$ level,
and other components at about the $10^{-21}$ level.
We estimate that about 60 components of coefficients for Lorentz and
CPT violation would be accessible at the Planck-scale.

\vspace{.4cm}\noindent
{\bf Rubidium Clocks}\\[.2cm]
Experiments with $^{87}$Rb are similar in many ways to ones
with $^{133}$Cs.
The clock transition $\ket{2,0} \rightarrow \ket{1,0}$,
is insensitive to Lorentz and CPT violation,
and so is a suitable reference signal.
A Zeeman hyperfine transition such as
$\ket{2,1} \rightarrow \ket{2,0}$
is a potential signal transition.
Like H and $^{133}$Cs, $^{87}$Rb
has an unpaired electron,
an is therefore sensitive to similar electron parameters
as discussed for those systems.
The sensitivity to proton parameters is also similar
to that for $^{133}$Cs, up to factors of order unity,
because the Schmidt nucleon for $^{87}$Rb
is a proton with angular momentum 3/2.
An advantage from the theoretical viewpoint
is the magic neutron number,
which aids in calculational reliability
and leads to cleaner results~\cite{kla}.
A considerable range of Lorentz and CPT bounds could
be envisaged for $^{87}$Rb with ideas along these lines.

\vspace{.4cm}\noindent
{\bf Other Spacecraft}\\[.2cm]
Lorentz and CPT tests could be done with
on a variety of space platforms.
Missions where the speeds of the craft with respect to the Sun
are larger than the speed $\be_s$ for Earth-orbiting satellites
are of particular interest.
One possibility is the proposed SpaceTime~\cite{spacetime} experiment,
which would attain $\be \simeq 10^{-3}$ on
a trajectory sweeping from Jupiter in towards the Sun.
This mission will fly
$^{111}$Cd$^+$, $^{199}$Hg$^+$, and $^{171}$Yb$^+$ ion clocks in a
craft rotating several times per minute.
This rotation rate would offer the possibility of
gathering data for a Lorentz and CPT test in
as little as 15 minutes.
The clock transitions
$\ket{1,0} \rightarrow \ket{0,0}$
are insensitive to Lorentz and CPT violation
for all three clocks,
and so could be used as reference signals.
Zeeman hyperfine transitions such as
$\ket{1,1} \rightarrow \ket{1,0}$
are sensitive to Lorentz- and CPT-violating effects
in the standard-model extension
and could provide signal clocks.
In the context of the Schmidt model,
all three clocks are sensitive to the neutron parameters
$\tib_3^n$, $\tid_3^n$, $\tig_d^n$ in the clock frame.
Such experiments would be of particular interest
because none of the above neutron parameters can be probed
with the proposed ISS experiments.
Several tests for Lorentz and CPT violation
would be possible by seeking variations
in the signal-clock outputs at
the spacecraft rotation frequency $\om_{ST}$ and also at
$2\om_{ST}$.
Experiments in this category would gain an order of magnitude
advantage over Earth-based or Earth-orbit experiments
because of their larger boost factors.

\vspace{.4cm}\noindent
{\bf 4. Discussion}\\[.2cm]
The standard-model extension is a microscopic theory
predicting possible minuscule Lorentz- and CPT-violating effects
in physical systems.
Some of the experimental challenges facing
measurements of such effects can be overcome by mounting experiments
on satellites orbiting the Earth.
In particular, atomic clocks planned for
the International Space Station
will be able to exploit the
relatively high rotation rates of the ISS as well as the
relatively high speed relative to the Earth to gain
sensitivity to relativistic effects within the context of the
standard-model extension.
Other experiments of interest in this context include
satellite-mounted microwave oscillators.

\vspace{.4cm}\noindent
{\bf 5. Acknowledgments}\\[.2cm]
I thank Robert Bluhm, Alan Kosteleck\'y, and Chuck
Lane for their collaboration on this work.
This work was partially supported by a grant from
Northern Michigan University.


\vspace{.4cm}\noindent
{\bf References}
\begin{thebibliography}{99}

\bibitem{spaceexpt}
R.\ Bluhm \etal, Phys.\ Rev.\ Lett.\ {\bf 88}, 090801 (2002).


\bibitem{ck}
D.\ Colladay and V.A.\ Kosteleck\'y, Phys.\ Rev.\ D {\bf 55}, 6760
(1997); {\bf 58}, 116002 (1998); Phys.\ Lett.\ B {\bf 511}, 209
(2001); V.A.\ Kosteleck\'y and R.\ Lehnert, Phys.\ Rev.\ D {\bf
63}, 065008 (2001).

\bibitem{cpt01}
For a broad overview of Lorentz and CPT symmetry issues,
see, for example,
V.A.\ Kosteleck\'y, ed., \it CPT and Lorentz Symmetry II, \rm
World Scientific, Singapore, 2002.

\bibitem{kps}
V.A.\ Kosteleck\'y and S.\ Samuel, Phys.\ Rev.\ D {\bf 39}, 683
(1989); {\bf 40}, 1886 (1989); Phys.\ Rev.\ Lett.\ {\bf 63}, 224
(1989); {\bf 66}, 1811 (1991); V.A.\ Kosteleck\'y and R.\ Potting,
Nucl.\ Phys.\ B {\bf 359}, 545 (1991); Phys.\ Lett.\ B {\bf 381},
89 (1996); Phys.\ Rev.\ D {\bf 63}, 046007 (2001); V.A.\
Kosteleck\'y, M.\ Perry, and R.\ Potting, Phys.\ Rev.\ Lett.\ {\bf
84}, 4541 (2000).

\bibitem{bek}
M.S.\ Berger and V.A.\ Kosteleck\'y, Phys.\ Rev.\ D {\bf 65},
091701(R) (2002).


\bibitem{chklo}
S.M.\ Carroll \etal, Phys.\ Rev.\ Lett.\ {\bf 87}, 141601 (2001);
Z.\ Guralnik \etal, Phys.\ Lett.\ B {\bf 517} 450 (2001);
A.\ Anisimov \etal, Phys.Rev. D {\bf 65}, 085032 (2002).

\bibitem{bckp}
O.\ Bertolami {\it et al.}, Phys.\ Lett.\ B {\bf 395}, 178 (1997).

\bibitem{kexpt}
KTeV Collaboration, Y.B.\ Hsiung {\it et al.}, Nucl.\ Phys.\
Proc.\ Suppl.\ {\bf 86}, 312 (2000).

\bibitem{ak}
V.A.\ Kosteleck\'y, Phys.\ Rev.\ Lett.\ {\bf 80}, 1818 (1998);
Phys.\ Rev.\ D {\bf 61}, 016002 (2000); {\bf 64}, 076001 (2001).

\bibitem{bexpt}
OPAL Collaboration, R.\ Ackerstaff {\it et al.}, Z.\ Phys.\ C {\bf
76}, 401 (1997); DELPHI Collaboration, M.\ Feindt {\it et al.},
preprint DELPHI 97-98 CONF 80 (1997); BELLE Collaboration, K.\ Abe
{\it et al.}, Phys.\ Rev.\ Lett.\ {\bf 86}, 3228 (2001).


\bibitem{ckpvi}
V.A.\ Kosteleck\'y and R.\ Potting, Phys.\ Rev.\ D {\bf 51}, 3923
(1995); D.\ Colladay and V.A.\ Kosteleck\'y, Phys.\ Lett.\ B {\bf
344}, 259 (1995); Phys.\ Rev.\ D {\bf 52}, 6224 (1995); V.A.\
Kosteleck\'y and R.\ Van Kooten, Phys.\ Rev.\ D {\bf 54}, 5585
(1996); N.\ Isgur \etal, Phys.\ Lett.\ B {\bf 515}, 333 (2001).

\bibitem{kr}
J.L.\ Rosner and S.A.\ Slezak,
Am.\ J.\ Phys.\ {\bf 69} 44 (2001);
V.A.\ Kosteleck\'y and A.\ Roberts,
Phys.\ Rev.\ D {\bf 63}, 096002 (2001).


\bibitem{cfj}
S.M. Carroll, G.B. Field, and R. Jackiw, Phys. Rev. D {\bf 41},
1231 (1990);
R.\ Jackiw and V.A.\ Kosteleck\'y, Phys.\ Rev.\ Lett.\ {\bf 82},
3572 (1999).

\bibitem{km}
V.A.\ Kosteleck\'y and M.\ Mewes, Phys.\ Rev.\ Lett.\ {\bf 87},
251304 (2001); preprint IUHET 449 (hep-ph/0205211).


\bibitem{vh}
V.W.\ Hughes {\it et al.}, Phys.\ Rev.\ Lett.\ {\bf 87}, 111804
(2001); R.\ Bluhm \etal, Phys.\ Rev.\ Lett.\ {\bf 84}, 1098
(2000).


\bibitem{eexpt}
H.\ Dehmelt {\it et al.}, Phys.\ Rev.\ Lett.\ {\bf 83}, 4694
(1999); R.\ Mittleman {\it et al.}, Phys.\ Rev.\ Lett.\ {\bf 83},
2116 (1999); G.\ Gabrielse {\it et al.}, Phys.\ Rev.\ Lett.\ {\bf
82}, 3198 (1999); R.\ Bluhm \etal, Phys.\ Rev.\ Lett.\ {\bf 82},
2254 (1999); Phys.\ Rev.\ Lett.\ {\bf 79}, 1432 (1997); Phys.\
Rev.\ D {\bf 57}, 3932 (1998).

\bibitem{eexpt2}
B.\ Heckel {\it et al.}, in B.N.\ Kursunoglu \etal, eds., \it
Elementary Particles and Gravitation, \rm Plenum, New York, 1999;
R.\ Bluhm and V.A.\ Kosteleck\'y, Phys.\ Rev.\ Lett.\ {\bf 84},
1381 (2000).

\bibitem{ccexpt}
V.W.\ Hughes, H.G.\ Robinson, and V.\ Beltran-Lopez, Phys.\ Rev.\
Lett.\ {\bf 4} (1960) 342; R.W.P.\ Drever, Philos.\ Mag.\ {\bf 6}
(1961) 683; J.D.\ Prestage {\it et al.}, Phys.\ Rev.\ Lett.\ {\bf
54} (1985) 2387; S.K.\ Lamoreaux {\it et al.}, Phys.\ Rev.\ A {\bf
39} (1989) 1082; T.E.\ Chupp {\it et al.}, Phys.\ Rev.\ Lett.\
{\bf 63} (1989) 1541.


\bibitem{lh}
C.J.\ Berglund {\it et al.}, Phys.\ Rev.\ Lett.\ {\bf 75} (1995)
1879; L.R.\ Hunter {\it et al.}, in V.A.\ Kosteleck\'y, ed., \it
CPT and Lorentz Symmetry, \rm World Scientific, Singapore, 1999.


\bibitem{db}
D.\ Bear {\it et al.}, Phys.\ Rev.\ Lett.\ {\bf 85}, 5038 (2000).


\bibitem{dp}
D.F.\ Phillips {\it et al.}, Phys.\ Rev.\ D {\bf 63}, 111101
(2001); M.A.\ Humphrey {\it et al.}, physics/0103068; Phys.\ Rev.\
A {\bf 62}, 063405 (2000).


\bibitem{kla}
V.A.\ Kosteleck\'y and C.D.\ Lane, Phys.\ Rev.\ D {\bf 60}, 116010
(1999); J.\ Math.\ Phys.\ {\bf 40}, 6245 (1999).

\bibitem{parcs}
N.\ Ashby, presented at the 2nd Pan Pacific Basin Workshop on
Microgravity Science, Pasadena, January 2001.


\bibitem{aces}
P.\ Laurent {\it et al.}, Eur.\ Phys.\ J.\ D {\bf 3} (1998) 201.


\bibitem{race}
C.\ Fertig {\it et al.}, presented at the Workshop on Fundamental
Physics in Space, Solvang, June 2000.


\bibitem{sumo}
S.\ Buchman {\it et al.}, Adv.\ Space Res.\ {\bf 25}, 1251 (2000).


\bibitem{spacetime}
L.\ Maleki and J.D.\ Prestage, in C.\ L\"ammerzahl {\it et al.},
eds., \it Testing Relativistic Gravity in Space: Gyroscopes,
Clocks, Interferometers, \rm Springer-Verlag, Berlin, 2001.


\end{thebibliography}
\end{document}